\documentclass[aps,twocolumn,prb]{revtex4-2}%
\usepackage{graphicx}
\usepackage{amssymb}
\usepackage{graphicx}
\usepackage{amsmath}
\usepackage{amsfonts}
\usepackage{color}
\usepackage[utf8]{inputenc}
\usepackage[T1]{fontenc}
\usepackage{mathptmx}
\usepackage{hyperref}
\usepackage{dcolumn}
\usepackage{bm}%
\providecommand{\U}[1]{\protect\rule{.1in}{.1in}}
\begin{document}
\title{Robust hybridization gap in a Kondo Insulator YbB${}_{12}$ probed by
femtosecond optical spectroscopy}
\author{A.R. Pokharel,$^{1}$ S.Y. Agustsson,$^{1}$ V.V. Kabanov,$^{2}$ F. Iga,$^{3}$
T. Takabatake,$^{4}$ H. Okamura$^{5}$ and J. Demsar$^{1}$}
\affiliation{$^{1}$Institute of Physics, Johannes Gutenberg-University Mainz, 51099 Mainz, Germany}
\affiliation{$^{2}$Jozef Stefan Institute, 1000 Ljubljana, Slovenia}
\affiliation{$^{3}$College of Science, Ibaraki University, Mito 310-8512, Japan}
\affiliation{$^{4}$Graduate School of Advanced Sciences of Matter, Hiroshima University, Japan}
\affiliation{$^{5}$Graduate School of Advanced Technology and Science, Tokushima
University, Japan}


\begin{abstract}
In heavy fermions the relaxation dynamics of photoexcited carriers has been
found to be governed by the low energy indirect gap, E$_{g}$, resulting from
hybridization between localized moments and conduction band electrons. Here,
carrier relaxation dynamics in a prototype Kondo insulator YbB${}_{12}$ is
studied over large range of temperatures and over three orders of magnitude.
We utilize the intrinsic non-linearity of dynamics to quantitatively determine
microscopic parameters, such as electron-hole recombination rate. The
extracted value reveals that hybridization is accompanied by a strong charge
transfer from localized 4f-levels. The results imply the presence of a
hybridization gap up to temperatures of the order of E$_{g}$/k$_{B}\approx200$
K, which is extremely robust against electronic excitation. Finally, below 20
K the data reveal changes in the low energy electronic structure, attributed
to short-range antiferromagnetic correlations between the localized levels.

\end{abstract}
\maketitle

\section{\bigskip Introduction}

The interaction between multiple degrees of freedom giving rise to the exotic
phases of matter is one of the most intriguing aspects of modern solid-state
physics. Heavy fermion compounds with partially filled 4\textit{f-} or
5\textit{f} -electron shells present one of the most challenging material
classes, hosting different electronic, magnetic and thermodynamic phases at
low temperatures (T) \cite{Stewart,Riseborough,Degiorgi}. At high-T, localized
\textit{f}-electrons are only weakly interacting with conduction (\textit{c})
electrons. As the temperature is lowered below the material specific Kondo
temperature (\textit{T${}_{K}$}), the localized moments, residing in the sea
of conduction electrons, hybridize with them. This \textit{c-f} hybridization
results in the formation of an (indirect) hybridization gap in the density of
states (DOS) \cite{Stewart,Coleman,Hewson}. In metallic heavy fermion systems
the Fermi level lies in one of the flat \textit{c-f} hybridized bands while in
Kondo insulators Fermi level resides within the hybridization gap. Upon
further cooling, inter-site correlations of localized magnetic moments in a
dense periodic Kondo lattice can give rise to a magnetic long range order.

The exact T-dependence of the low-energy electronic structure, which manifests
itself e.g. in a crossover between the small and large Fermi surface in heavy
fermions \cite{Kotliar,Guettler} and in the T-dependence of the electronic
effective mass, as well as the underlying microscopic mechanisms, have been
the topic of intense research efforts in the recent years
\cite{Kotliar,Guettler,Ernst,Elmers}.

In addition to photoemission spectroscopy (PES) \cite{Guettler,Elmers} and
tunneling spectroscopy \cite{Ernst}, femtosecond real-time approaches have
also been shown to be highly sensitive to changes in the low-energy electronic
structure of heavy electron systems. Here, studies on a series of heavy
fermions
\cite{DemsarHF,DemsarHFhybGap,DemsarHFreview,Burch,YbAl3,Chia,ZXShen,Liu} as
well as on the Kondo insulator SmB${}_{6}$
\cite{DemsarHFhybGap,DemsarHFreview,RickSmB6} were performed, demonstrating
the relaxation of photoexcited carriers to be governed by the presence of the
hybridization gap near the Fermi level \cite{DemsarHFreview}. The dramatic
slowing down of carrier relaxation at low temperatures
\cite{DemsarHF,DemsarHFhybGap,DemsarHFreview,Burch,YbAl3,Chia,Liu}, in some
cases by orders of magnitude \cite{DemsarHFreview}, was qualitatively
accounted for by the phenomenological model \cite{DemsarHFreview}, originally
developed to describe the relaxation dynamics in fully-gaped superconductors
\cite{RT,VVKRT,MgB2,NbN,DemsarSCrev}. The model, given by the set of coupled
non-linear rate equations, describes the recombination of photoexcited
electron-hole pairs across the indirect hybridization gap via emission of
large-momentum, high-frequency phonons. The re-absorption of the latter
eventually limits the relaxation process. In addition to time-resolved studies
with near-infrared pulses
\cite{DemsarHF,DemsarHFhybGap,DemsarHFreview,Burch,YbAl3,Chia,Liu}, the
dynamics of Kondo insulators was recently investigated also by time-resolved
PES (tr-PES) \cite{Okawa} and transient THz spectroscopy \cite{RickSmB6},
providing direct spectroscopic support to the above scenario. While there is
mounting evidence supporting the above description, the non-linear nature of
the relaxation processes should enable quantifying the underlying microscopic
parameters and provide details on the nature of the \textit{c-f}
hybridization. For such a quantitative characterization, a well studied
prototypical Kondo insulator should be used.

YbB${}_{12}$ is one of the most studied Kondo insulators
\cite{Okamura,Takeda,Hagiwara,Iga,Okawa}, that recently regained interest also
as a candidate for hosting non-trivial topological surface state
\cite{Xiang,Xu,Tsvelik,Peters}. Optical data reveal a fully open indirect
hybridization gap of $E{}_{g}\sim20$ meV below $\sim25$ K, with a crossover to
the metallic behavior near $\sim80$ K \cite{Okamura}. Similar values of
E${}_{g}$ at low-T were extracted also from the high resolution PES
\cite{Takeda,Hagiwara}, while transport measurements suggest $E{}_{g}\sim13$
meV \cite{Iga}. Interestingly, a combined PES and tr-PES study suggests the
gap closing above $T\approx100$ K \cite{Okawa}. Thus, while the existence of a
gap at low-T is unambiguous, its $T$-dependence and robustness against
external stimuli has yet to be clarified.

In this paper, we explore in detail the temperature and excitation density
dependence of carrier relaxation dynamics in one of the best-known Kondo
insulators YbB${}_{12}$. We demonstrate that the (non-linear) dynamics over
large range of temperatures and over three orders of magnitude in excitation
density can be quantitatively described by the relaxation bottleneck model
\cite{DemsarHFhybGap,DemsarHFreview}. Quantitative analysis provides access to
the microscopic electron-hole recombination rate, $R$, which is found to be
about 3 orders of magnitude lower than in superconductors with comparable gap
energy \cite{MgB2,NbN,DemsarSCrev}. We ascribe this observation to an enhanced
density of states, which is a result of the charge transfer from localized 4f
states that is accompanying hybridization. We show that the indirect
hybridization gap, $E{}_{g}\sim15$ meV, persists near to the room temperature
and is extremely robust against electronic excitation, up to the absorbed
energy densities of $\approx140$ meV per unit cell volume (ucv). Finally, for
$T\lesssim20$ K, the data suggest changes in the low energy electronic
structure, likely due to short range antiferromagnetic correlations between
local moments.

\section{Experimental details}

Here, we study temperature and excitation density dependent carrier relaxation
dynamics by tracking reflectivity changes of YbB${}_{12}$ single crystal using
60 femtosecond (fs) near-infrared optical pulses (800 nm) for both pump and
probe. A disk-shaped sample, 4.5 mm in diameter, was cut from the single
crystal \cite{Iga} and mechanically polished for optical measurements
\cite{Okamura}. Experiments were performed with a Ti:sapphire laser amplifier
operating at 250 kHz, employing a double modulation fast-scan technique.

The reported measurements cover a range of excitation densities between 0.5 $\mu$%
J/cm$^{2}$ and 0.5 mJ/cm$^{2}$. Thus, continuous laser heating effects should be considered. The average heating of the excited
spot is mainly governed by the thermal conductivity of the sample and can be
easily calculated using a simple steady-state heat diffusion model
\cite{MihailovicDemsar}. Thermal conductivity of YbB$_{12}$ exceeds 10 W/mK
over the entire range of temperatures in question \cite{Kappa}. Taking into
account the optical constants \cite{Okamura}, and considering the worst case
scenario (base temperature of 5 K and excitation density of 410 $\mu$%
J/cm$^{2}$), the average laser heating amounts to only 1.1 degrees in
YbB$_{12}$ and can thus be neglected.

\section{Results}

Figure 1 presents the raw data recorded as a function of (a) excitation
density (at base temperature of 5 K) and (b) temperature (at a constant
fluence, $F=8$ $\mu$J/cm${}^{2}$). For $F\lesssim20$ $\mu$J/cm${}^{2}$, the
excitation density dependence data display slightly sub-linear dependence of
amplitude on fluence, yet the characteristic timescales remain constant. For
higher fluences, both, amplitudes as well as the characteristic timescales,
show a strong dependence on F. To follow the T-dependence of the dynamics in
the low excitation limit we thus chose $F=8$ $\mu$J/cm${}^{2}$ (Fig. 1b).

\begin{figure}[h]
\centerline{\includegraphics[width=90mm]{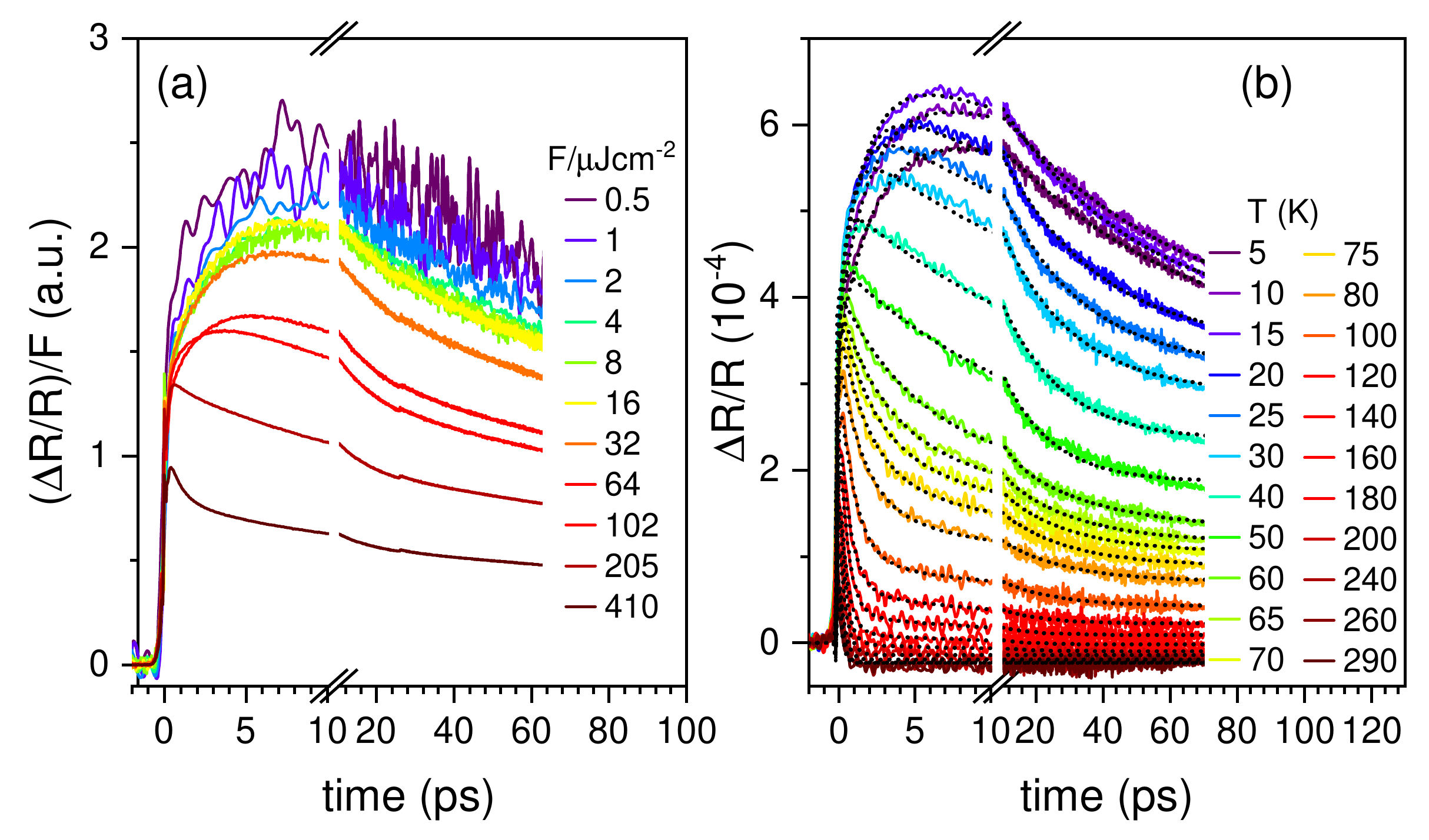}}\caption{(color online):
Dynamics of photoinduced change in reflectivity of YbB${}_{12}$ single crystal
at 800 nm. (a) Dynamics recorded at 5 K base temperature as a function of F
over three orders of magnitude. All traces were normalized to F. (b)
T-dependence of reflectivity traces recorded at F = 8 $\mu$J/cm${}^{2}$.
Dashed lines are fits to the data (see text). Note the break in the x-axis.
Continuous laser heating is less than 1 K over the entire range of
temperatures and excitation densities.}%
\end{figure}

\subsection{T-dependent dynamics in the weak perturbation regime}

We start by discussing the T-dependence of dynamics in the low excitation
regime. From Fig. 1(b) one can clearly see a strong T-dependence of the
recovery time, which varies between several tens of ps at 5 K to about 1 ps at
80 K. Moreover, the data at $T\leq40$ K display a slow picosecond buildup.
While such was not observed in SmB${}_{6}$ \cite{DemsarHFhybGap}, in
superconductors it has been attributed to pair breaking by absorption of
high-frequency phonons, created during the relaxation of hot carriers towards
the gap \cite{MgB2,NbN}.

To quantify the T-dependence of the amplitude and recovery time we first fit
the photo-induced reflectivity traces with
\begin{equation}
\frac{{\Delta}R}{R}(t)=H\left(  t\right)  [A_{1}(1-{\mathrm{e}}^{-\frac
{t}{t_{rise}}})\ e^{-\gamma t}+A_{2}].\label{SimpleFit}%
\end{equation}
Here A${}_{1}$, t${}_{rise}$ and $\gamma$ are the amplitude, rise time and the
decay rate of the response, A${}_{2}$ accounts for a remaining bolometric
signal at $t\gg100$ ps, and $H\left(  t\right)  $ is the step function with
resolution limited rise time of $\mathrm{\sim}$ 100 fs. This simple fit
function accounts well the data at $T<60$ K; at higher-T the entire recovery
time-window is better fit by a double exponential decay.

The overall slowing down of the relaxation upon cooling is an indication of an
energy gap in the excitation spectrum. It has been observed both in
superconductors \cite{VVKRT,MgB2,NbN,DemsarSCrev} as well as in heavy electron
systems \cite{DemsarHF,DemsarHFhybGap,DemsarHFreview,YbAl3,RickSmB6},
attributed to a boson (phonon) bottleneck \cite{DemsarHFreview,VVKRT}. Here,
following photoexcitation, hot electrons and holes first relax via e-e and
e-ph collisions towards the gap, resulting in excess densities of
electrons-hole pairs (EHP) and high frequency ($\hbar\omega>E{}_{g}$) phonons
(HFP). We assume the electron-hole symmetry with densities of electrons and
holes being the same ($n$). The presence of the gap hinders the recombination
of excess EHPs due to the competing creation of pairs by HFP absorption,
resulting in a phonon bottleneck \cite{VVKRT}. In this case, it is the slow
decay of the HFP population (via anharmonic decay or via diffusion out of the
excitation volume) that governs the relaxation rate $\gamma$ of the coupled
EHP-HFP system back to equilibrium.

The time evolution of the system is given by a set of two coupled differential
equations \cite{VVKRT}:%

\begin{align}
\dot{n}  &  =\beta N-Rn^{2}+n_{0}\delta(t)\nonumber\\
\dot{N}  &  =\ -\left[  \beta N-Rn^{2}\right]  /2-\left(  N-N_{0}\right)
{\tau}_{\gamma}^{-1}+N_{0}\delta(t) \label{Eq2}%
\end{align}

Here, $n$\textit{ }and $N$ are the EHP and HFP densities, respectively,
$\beta$ is the probability of pair creation by HFP absorption and \textit{R}
is the bare EHP recombination rate with the creation of HFP. ${\tau}_{\gamma
}^{-1}$ is the decay rate of HFP, governed, \textit{e.g.} by anharmonicity,
and can be approximated as being T-independent. Note that the microscopic
parameter ${\tau}_{\gamma}$ differs from the measured recovery rate $\gamma$;
the latter reflecting the recovery of the coupled EHP-HFP system \cite{VVKRT}.
Finally, $n{}_{0}$ and $N{}_{0}$ are the EHP and HFP densities created during
the initial avalanche process (on a 100-fs timescale). These are determined
from the absorbed energy density, $\Theta$, considering an energy E${}_{g}$
per HFP and E${}_{g}$/2 per electron and hole, such that $n_{0}=\frac{2\Theta
r}{E_{g}}$ and $N_{0}=\frac{\Theta\left(  1-r\right)  }{E_{g}}$. Here $0\leq
r\leq1$ is the fraction of $\Theta$ in the EHP channel, and can be determined
from the F-dependence of the ps build-up, addressed in Section \ref{Fluence}.

We now make use of the bottleneck model \cite{DemsarHFhybGap,VVKRT} for a
quantitative analysis of the data to gain information on the T-dependence of
the low energy gap in YbB${}_{12}$. We analyze the $T$-dependence of the
amplitude of reflectivity transient and its relaxation rate (the $T$- and
$F$-dependence of the ps build-up is addressed below). As ${\tau}_{\gamma}$ is
the largest timescale, the initial thermalization between EHPs and HFPs on the
ps timescale results in a quasi-equilibrium state, given by the detailed
balance equation
\begin{equation}
Rn_{s\ }^{2}=\beta N_{s},\label{detailedBalance}%
\end{equation}
where $n{}_{s}$ and $N{}_{s}$ are the quasi-thermal concentrations of EHPs and
HFPs at some new effective temperature $T{}^{\ast}$\cite{Parker} - see also
Appendix A. Assuming no energy is yet transferred to phonons with energy
smaller than the gap, this quasi-stationary case is given by
\begin{equation}
Rn_{s}^{2}=\beta N_{s}\ ;n_{s}=\frac{\beta}{4R}\left[  \sqrt{1+8R\beta
^{-1}\left(  2n_{0}+N_{0}\right)  \ \ }-1\right]  .\label{Eq4}%
\end{equation}
Once this quasi-equilibrium state has been reached, the coupled system relaxes
through the decay of HFPs.

In the weak perturbation limit, the amplitude or reflectivity change,
\textit{A${}_{1}$}, is proportional to excess EHP density. At finite
temperatures, $A_{1}\ \mathrm{\propto(}n_{s}-n_{T})$, where $n_{T}$ is the
density of thermally excited EHPs. Taking into account the detailed balance
equation, Eq.(\ref{Eq4}), it has been shown
\cite{DemsarHFhybGap,VVKRT,RickSmB6,Liu}, that $n_{T}(T)$ and thus $E_{g}(T)$
can be extracted from the measured \textit{A}${}_{1}(T)$ via
\cite{DemsarHFreview,VVKRT}:
\begin{equation}
n_{T}\left(  T\right)  \propto A_{1}\left(  0\right)  A_{1}^{-1}\left(
T\right)  -1.\label{Eq5}%
\end{equation}
Here $A_{1}\left(  0\right)  $ is the amplitude in the limit when
$T\rightarrow0$ K. With, $n_{T}\left(  T\right)  \propto N(0){\left(
E_{g}T\right)  }^{p}e^{-E_{g}/2k_{B}T}$, where $N(0)$ is the normal state
density of states and power $p$ depends on the exact shape of the DOS,
$E_{g}(T)$ can be estimated.

Inset to Figure 2(a) presents the $T$-dependence of $A{}_{1}$, obtained by
taking maximum values of reflectivity transients presented in Figure 1(b).
Upon cooling down, $A{}_{1}$ displays a continuous growth down to
$\mathrm{\approx}$ 20 K. Below 20 K, however, $A{}_{1}$ does not saturate, but
shows a decrease - similarly as in SmB${}_{6}$ \cite{RickSmB6}. The main panel
of Fig. 2(a) presents $n_{T}(T)$, extracted through Eq.(\ref{Eq5}), where the
maximum amplitude was used as $A{}_{1}(0)$. Above 20 K, we fit the extracted
$n_{T}(T)$ using $p=1/2$, which corresponds to the variation of the DOS near
the gap edge, similar to a superconductor (the fit is rather insensitive to
small variations in $p$). The gradual increase of $n_{T}$ with increasing T
suggest that the hybridization gap is present to $T>200$ K. In the fit, we
consider $E_{g}$ to be T-independent within the temperature range in question
($T<200$ K). We did consider different T-dependences of $E_{g}$, similar to
those extracted for Ce${}_{3}$Bi${}_{4}$Pt${}_{3}$ \cite{Riseborough,Fisk},
yet no major improvement in the fit quality can be achieved. The extracted
values for $E_{g}$, shown in Figure 2(a), are around 15 meV, consistent with
earlier spectroscopic studies \cite{Iga,Okawa}.

We note a decrease in amplitude below $\approx20$ K, reproduced by several
measurements on different days. Since at these low temperatures $n_{T}$ is
always low compared to the density of photoexcited EHPs, a decrease in the
signal is most likely caused by a (slight) increase in the gap energy scale,
or by sharpening of the peaks in the DOS, as suggested by recent tunneling
data on SmB${}_{6}$ \cite{Paglione}.

\begin{figure}[h]
\centerline{\includegraphics[width=90mm]{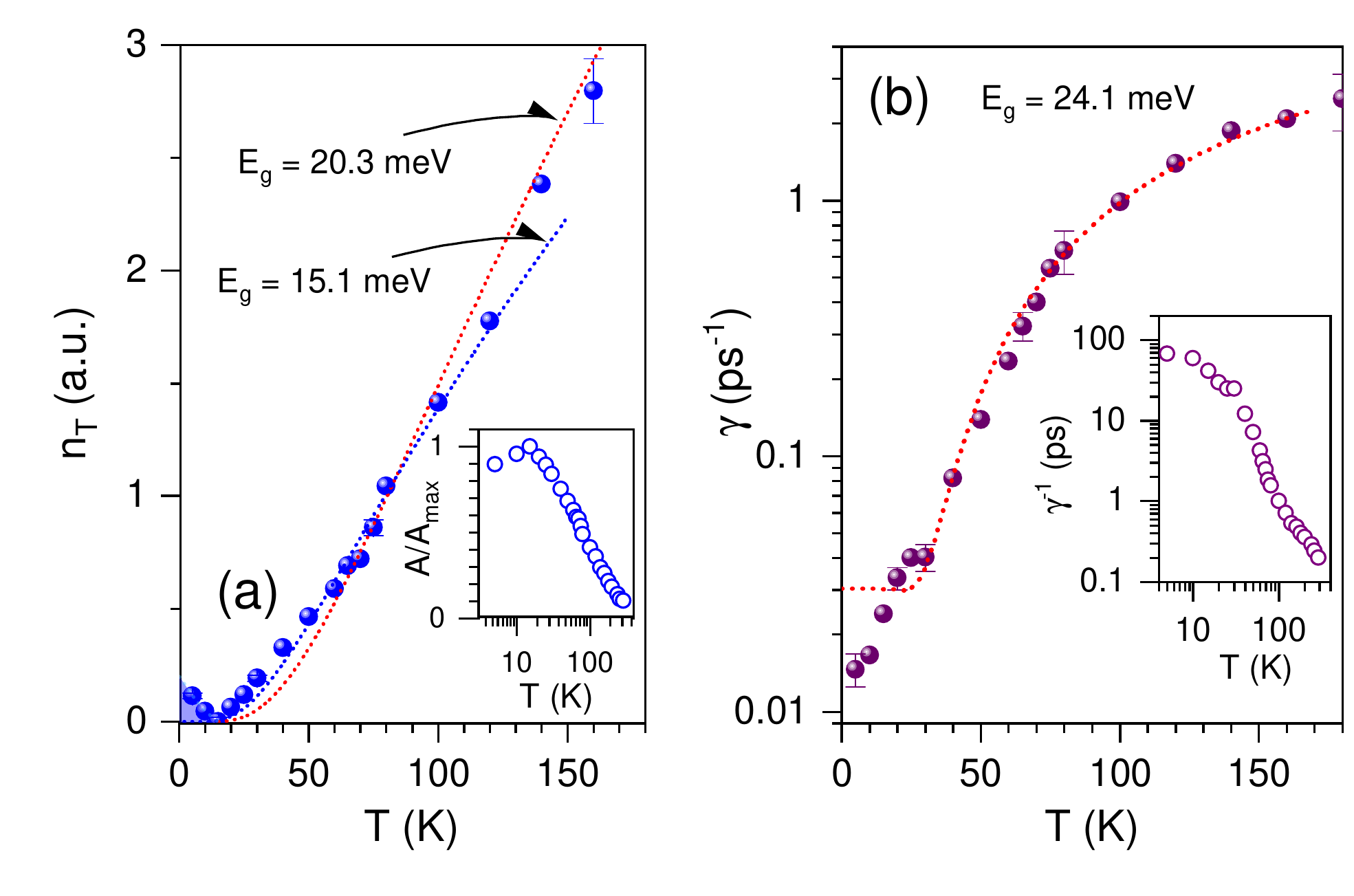}}\caption{(color online):
T-dependence of (a) thermally excited EHP density extracted from the
T-dependence of $A_{1}$, presented in the inset, and (b) the relaxation rate
$\gamma$. The data were recorded at F = 8 $\mu$J/cm${}^{2}$. The solid spheres
are the experimental data and the dotted lines are fits to the data (see
text).}%
\label{fig2}%
\end{figure}

In the low excitation limit the recovery rate $\gamma$ has also been shown to
be governed by $n_{T}$ \cite{DemsarHFreview}, where $\gamma\propto\ [D{\left(
En_{T}+1\right)  }^{-1}+2n_{T}]$, with \textit{D} and \textit{E} being the
T-independent proportionality constants. Indeed, as shown in Figure 2(b),
$\gamma$ continuous to increase up to the highest T. Fitting the data, using
the same functional form for $n_{T}$, we obtain somewhat higher value for
$E{}_{g}=24$ meV. Also here, a departure from the high temperature behavior is
seen below $\approx20$ K, consistent with changes in the low energy structure.
Given the simple approximations for the shape of the DOS and assuming the gap
to be simply T-independent, the overall agreement with the model is very good.

The observation of departure from constant gap behavior below 20 K implies
changes in the low energy excitation spectrum at low-T. Similar departure from
the high-T behavior was observed at low-T also in SmB${}_{6}$ \cite{RickSmB6}.
Together with the large residual THz conductivity, the observation in
SmB${}_{6}$ was attributed to a (topological) surface state. In YbB${}_{12}$,
however, the residual THz optical conductivity is at least two orders of
magnitude lower than in SmB${}_{6}$ \cite{Okamura}, thus we exclude this
possibility. Instead, we argue that correlations between local moments give
rise to changes in the low energy excitation spectra. Indeed, inelastic
neutron scattering data do provide evidence for short range antiferromagnetic
fluctuations at comparable temperatures \cite{Mignot}.

\subsection{Excitation density dependence at 5 K\label{Fluence}}

The $T$-dependent study implies the hybridization to be present nearly up to
room temperature, where the relaxation time becomes comparable to the e-ph
thermalization in the metallic state. The question is, how robust is
\textit{c-f} \ hybridization with respect to the electronic excitation.
Moreover, the peculiar behavior of a finite, $T$- and $F$-dependent rise-time
needs addressing. As shown in Fig. 1, the instantaneous build-up of
photo-excited carrier density on a $\mathrm{\sim}$ 100 fs timescale is
followed by a further increase on a ps timescale. The delayed rise-time
gradually decreases with increasing excitation density and finally becomes
resolution limited for $F>100$ $\mu$J/cm${}^{2}$.

In superconductors, the observation of the ps buildup has been attributed to
Cooper pair-breaking by re-absorption of HFPs, taking place during the
formation of the quasi-stationary state between the condensate, broken pairs
and HFPs \cite{MgB2,NbN,PCCO}- see also Eq.(\ref{Eq4}). The non-linear nature
of this so-called pre-bottleneck dynamics \cite{RT,VVKRT} could be used
to determine microscopic parameters, such as the e-ph coupling constant in NbN
\cite{NbN}. Moreover, as such nonlinear dynamics is limited to range of
excitation densities, for which the resulting gap suppression is perturbative
only \cite{NbN}, such study provides means to study the gaps robustness
against electronic excitation.

For early time-delays, where the HFP decay term can be neglected, the Eqs.
(\ref{Eq2}) have analytic solutions for $n(t)$ and $N(t)$ - see Appendix A and
Refs. \cite{VVKRT}. Moreover, densities of thermally excited EHPs and
HFPs can also be neglected at low temperatures. The solutions thus depend only
on the microscopic parameters $R$, $\beta$ and the fraction of the absorbed
energy density, $\Theta$, in the EHP channel $0\leq r\leq1$. $\Theta$ is
determined via $\Theta=F(1-\mathcal{R})/\lambda_{opt}$, with the optical
penetration depth $\lambda_{opt}=$ 50 nm and reflectivity $\mathcal{R}=0.3$
extracted from optical data \cite{Okamura}.

\begin{figure}[h]
\centerline{\includegraphics[width=90mm]{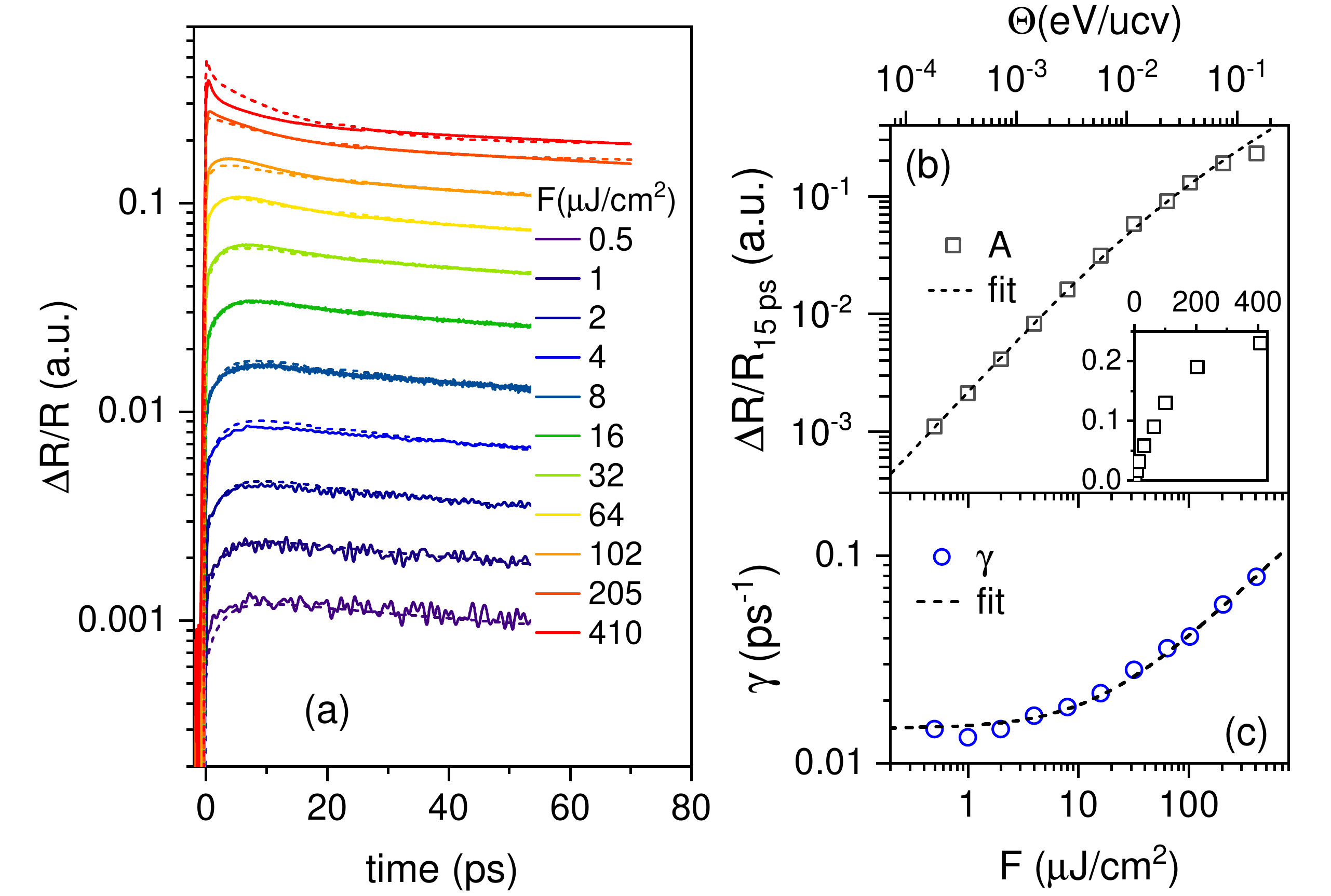}}\caption{ (color
online): The analysis of the F-dependence of transient reflectivity at 5 K
using the phonon-bottleneck model \cite{VVKRT}. Panel (a) presents the data
recorded for F spanning three orders of magnitude (solid lines) fit by the
global fit (dashed lines), providing access to parameters R, $\beta$ and r.
(b) The F-dependence of reflectivity change at 15 ps time delay (open black
squares), together with the fit (dashed line) with n${}_{s}$(F), given by Eq.
(\ref{Eq4}). Insert present the same data on the linear scale. (c) The
F-dependence of the relaxation rate $\gamma$ (open blue circles), fit by
\textit{$\gamma\propto(n_{s}+E)$} (dashed line). For all fits the same values
of microscopic parameters were used: r = 0.47 +/- 0.01, $\beta$ = 0.61 +/-
0.01 ps${}^{-1}$ and R = 0.14 +/- 0.01 ps${}^{-1}$ucv.}%
\label{fig4}%
\end{figure}

We perform a global fit to the $F$-dependent data, with the excitation density
$F$ spanning over three orders of magnitude. We first determine the
($F$-dependent) values of the recovery rate $\gamma$, by fitting the dynamics
for $t>10$ ps by $Be^{-\gamma t}+C$. The resulting function is then multiplied
by an analytic solution for pre-bottleneck kinetics - Eq.(\ref{eqref}) in
Appendix A. Figure 3(a) presents the global fit of the data, using $\gamma(F)$
shown in panel (c) and $E{}_{g}=15$ meV. An excellent agreement with the data
is obtained, especially considering that the large span of excitation
densities. The extracted global parameters are: $r=0.47+/-0.01$,
$\beta=0.61+/-0.01$ ps${}^{-1}$ and $R=0.14+/-0.01$ ps${}^{-1}$ucv. Note a
peculiar change in the character of the pre-bottleneck dynamics from being
governed by pair-generation (build-up) to being governed by the bi-molecular
recombination (decay) near 100 $\mu$J/cm$^{2}$ - see also simulation in
Appendix A. In superconductors such a change in dynamics has never been
observed, since quenching of the gap takes place before this regime can be
reached. The agreement between the data and the model thus implies no
pronounced reduction of E${}_{g}$ up to $F=400$ $\mu$J/cm$^{2}$, which
corresponds to $\Theta\approx140$ meV per unit cell (i.e. $\mathrm{\sim}$ 35
meV per Yb). For comparison, this value exceeds values in superconductors with
comparable gap sizes by two to three orders of magnitude \cite{DemsarSCrev}.

Figure 3(b) presents the F-dependence of $\Delta R/R\left(  15\text{
ps}\right)  $, where a quasi-stationary state, given by Eq.(\ref{Eq4}), is
established. The best fit with Eq.(\ref{Eq4}) is shown by the dashed line,
again in-line with the model. 

Finally, Figure 3(c) presents $\gamma($F$)$. For low temperatures, $\gamma$
has been shown to follow $\gamma\propto(n_{\mathrm{s}}+n_{\mathrm{T}})$
\cite{VVKRT}. Strictly speaking, this implies $\gamma{}^{-1}\rightarrow\infty$
in the limit of $T$, $F\rightarrow0$. However, the model does not take into
account extrinsic effects, as for example the diffusion of hot carriers out of
the probed volume \cite{YbAl3}. Indeed, both T- and F-dependent data show a
saturation of the relaxation time, \textit{i.e.}, $\gamma{}^{-1}%
(T,F\mathrm{\rightarrow}0)\approx60$ ps, which is likely limited by the
carrier transport into the bulk of the crystal. Taking this into account, we
fit $\gamma\propto(n_{\mathrm{s}}+E)$, with $E$ being a constant - see dashed
line in Figure 3(c). The agreement with the model over large range of
excitations further evidences the robustness of hybridization against
electronic excitation.

\section{Discussion}

Let us now turn to the extracted microscopic parameters. The ratio
$r\approx0.5$ implies a high \textit{e-ph} relaxation rate for hot carriers;
similar value is obtained in a conventional superconductor NbN \cite{NbN}. The
rate of the EHP creation by absorption of a phonon, $\beta$, is also similar
to values obtained in NbN \cite{NbN}. However, the extracted value of
recombination rate $R$ is in YbB${}_{12}$ about three orders of magnitude
lower than in NbN \cite{NbN}. To address this, let us consider the detailed
balance equation. From $\eta/R=$ $n_{T}^{2}/N_{T}$ it follows that the ratio
$\eta/R$ is governed by the (high temperature) densities of states of
electrons and phonons. $N_{T}$ can be estimated in the Debye approximation as
$N_{T}=9\nu E_{g}^{2}k_{B}T\omega_{D}^{-3}\exp(-E_{g}/k_{B}T)$, where $\nu$ is
the number of atoms per unit cell and $\omega_{D}$ is the Debye energy
\cite{VVKRT}. With $n_{T}\left(  T\right)  \simeq$ $N(0)\sqrt{\pi E_{g}k_{B}%
T}\exp(-E_{g}/2k_{B}T)$ it follows $\eta/R=\frac{N(0)^{2}\pi\omega_{D}^{3}%
}{9\nu E_{g}}$ \cite{VVKRT}. Clearly, only a variation in $N(0)$ can account
for a 1000-fold increase of $\eta/R$ in YbB$_{12}$ as compared to NbN. As the
density of states of conduction band electrons are comparable for the two
systems \cite{DOSNbN,DOSLuB12}, this implies that hybridization not only
results in the gap in the density of states, but is accompanied by a
pronounced charge transfer from the localized 4f states into the hybridized band.

\section{Conclusions}

In summary, we show that in YbB${}_{12}$ the hybridization gap persists to
temperatures of the order of $E_{g}/k_{B}$ and is extremely robust against
electronic excitation. This is likely applicable to a larger class of Kondo
lattice systems, as similar robustness can be inferred also from studies on
SmB$_{6}$ \cite{DemsarHFhybGap}, and heavy fermion systems
\cite{DemsarHFhybGap,ZXShen}. A quantitative analysis of the density dependent
carrier dynamics presented here, provides access to microscopic parameters,
revealing that hybridization is accompanied by a strong charge transfer from
the localized 4f levels. Finally, near 20 K, the relaxation dynamics in
YbB$_{12}$ show a departure from the high-T behavior, suggesting changes in
the low-energy gap structure that can be attributed to short-range
antiferromagnetic correlations between local moments.

\begin{acknowledgments}
This work was supported by the DFG in the framework of the Collaborative
Research Centre TRR173 268565370 (Project A05) and TRR288 422213477 (Project B08).
\end{acknowledgments}

\appendix*
\section{Simulations of pre-bottleneck dynamics}

Considering $\tau{_{\gamma}}$ being the largest timescale of the problem, the
early timescale dynamics describes the buildup of the quasi-equilibrium
between the EHP and HFP subsystems, \textit{i.e.} their densities reach a
quasi-thermal equilibrium (the effective temperature however differs from the
temperature of low frequency phonons). For this limiting case, exact
analytical solutions for the time-evolution of $n$ and $N$ have beed derived
for excitation densities where the resulting gap suppression is not too strong
\cite{MgB2,VVKRT}. At low temperatures the density of thermally excited EHPs
and HFPs can be neglected. Considering $\tau{_{\gamma}\rightarrow\infty}$ and
the initial conditions ($n(t\approx0)=n{}_{0}$ and $N(t\approx0)=N{}_{0}$),
the coupled differential equations, Eq. (\ref{Eq2}), have the following
analytical solution for $n(t)$:%

\begin{equation}
n\left(  t\right)  =\ \frac{\beta}{R}\ \left[  -\frac{1}{4}-\frac{1}{2\tau
}+\frac{1}{\tau\left(  1-Ke^{-\frac{t\beta}{\tau}}\right)  }\ \right]
\label{eqref}%
\end{equation}

Here, ${\tau}^{-1}$ and $K$ are the dimensionless parameters determined by the
initial conditions, $n{}_{0}$, $N{}_{0}$ and the microscopic constants $R$ and
${\beta}$ \cite{MgB2,VVKRT}:
\begin{align}
\tau{^{-1}}  & =\sqrt{\frac{1}{4}+\frac{2R}{\beta}(n_{0}+2N_{0})}\text{ \ }\\
\text{ \ \ }K  & =\frac{\frac{\tau}{2}(\frac{4Rn_{0}}{\beta}+1)-1}{\frac{\tau
}{2}(\frac{4Rn_{0}}{\beta}+1)+1}\nonumber
\end{align}

Figure 4 presents different limiting cases for $n(t)$ in this pre-bottleneck
regime, which depend on the microscopic parameters $R$ and $\beta$, as well as
on the initial conditions, which are given by the ratio $r$ and absorbed
energy density $\Theta$ \cite{DemsarSCrev}. Here, $n(t)$ is obtained by
convoluting Eq. (\eqref{Eq2}) with the Heaviside step-function, whose rise
time is chosen to be 100 fs, and reflects the time resolution of the
experiment as well as the characteristic time for the initial e-e and e-ph processes.

\begin{figure}[h]
\centerline{\includegraphics[width=90mm]{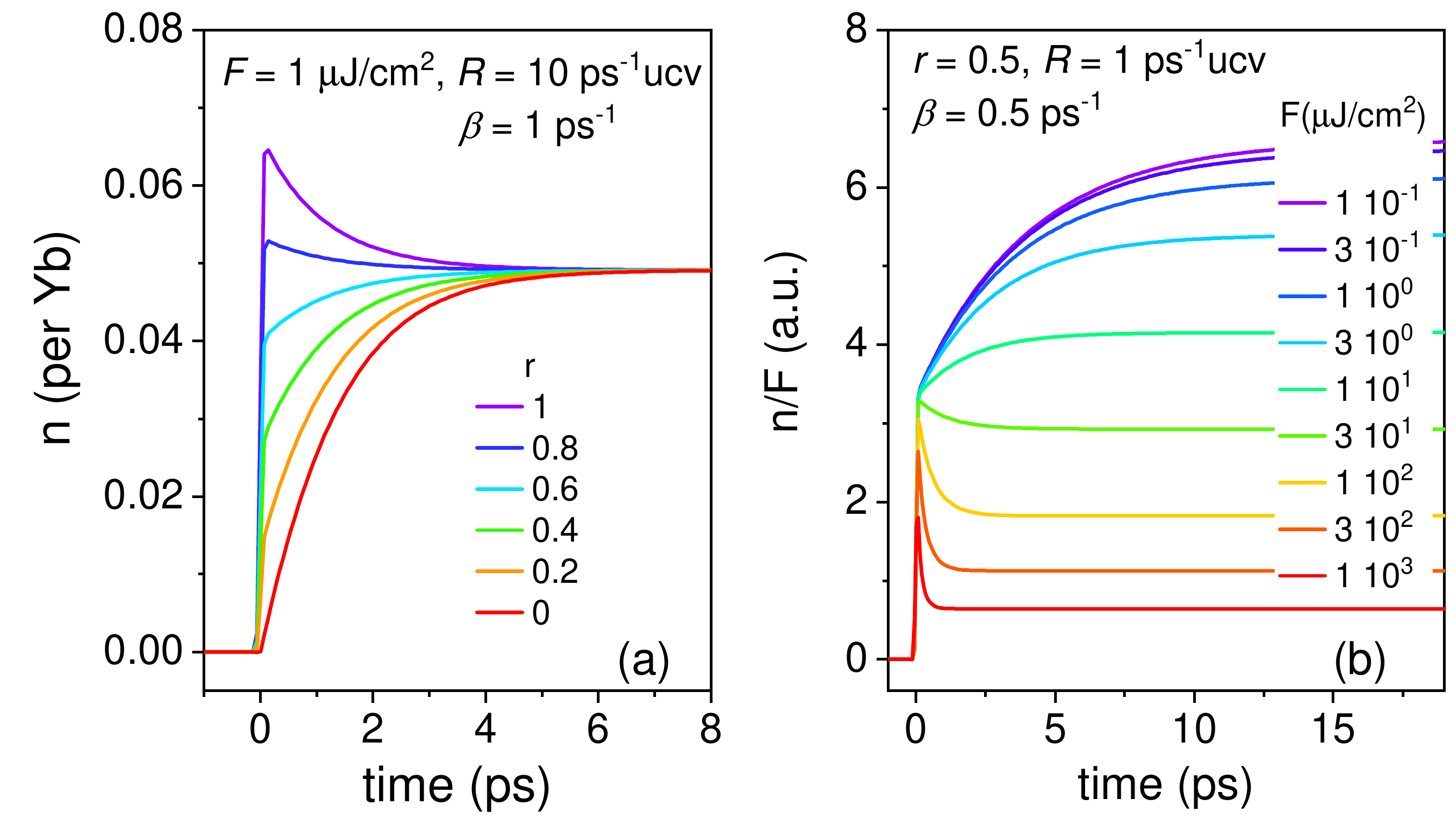}}\caption{ (color
online): Simulation of early stage QP dynamics in the absence of decay channel
(which proceeds via phonon escape/anharmonic decay). Panel (a) presents the
evolution for constant F = 1 mJ/cm${}^{2}$ and different values of r from 1
(all the energy initially in the QP channel) to 0 (all the energy initially in
the HFP channel). (b) Fluence dependence of dynamics for constant R, b, r (the
values are given in the plot). Here n(t) is normalized to excitation fluence,
F, to emphasize the nonlinearity of dynamics.}%
\label{S1}%
\end{figure}

In this simulation, we consider parameters relevant for YbB$_{12}$,
with E${}_{g}$ = 15 meV and excitation densities comparable to those used in
the experiment. From the optical conductivity data \cite{Okamura} at 1.55 eV
we extract the values of the dieletric functions $\epsilon{_{1}}$ $\approx$
-0.498 and $\epsilon{_{2}}$ $\approx$ 3.12, and determine the optical
penetration depth to be $\lambda_{opt}\approx50$ nm. The absorbed energy
density ($\Theta$), is determined from the optical penetration depth and
reflectivity at 1.55 eV of $\mathcal{R}=0.3$ as
\[
\Theta=\frac{F\left(  1-\mathcal{R}\right)  }{\lambda_{opt}}.
\]
As a reference, the incoming fluence $F=1$ mJ/cm${}^{2}$ corresponds to
${\Theta}=364$ meV/ucv or 91 meV/Yb.

Figure 4(a) presents the simulation of the time evolution of the EHP
density for constant excitation density $\Theta$ but different values of $r$.
Here, $r=1$ corresponds to the limit where all of the absorbed energy is in
initially in the EHP channel while $r=0$ corresponds to the other extreme
case, where the entire energy is initially transferred to HFPs. For all values
of $r$ the same quasi-equilibrium concentration of EHPs, $n_{s}$, is reached
(determined simply by the detailed balance equation, $Rn_{s}^{2}=\beta N_{s}%
$). However, over large range of values of parameter $r$, the initial
excitation is followed by the generation of additional EHPs via HFP
absorption, as manifested by the delayed buildup of $n_{s}$ in Figure 4(a).

Figure 4(b), on the other hand, presents the excitation density dependence of
EHP dynamics using constant values of parameters $R$, $\beta$ and $r$. It is
reasonable to assume these parameters are in the first approximation
independent on excitation density. The absorbed energy density $\Theta$ (or
$F$) is varied here over four orders of magnitude. At the lowest fluences, the
dynamics is obviously linear, governed solely by the electron-hole pair
generation rate $\beta$. For higher excitation densities the dynamics becomes
nonlinear, reflecting the bi-molecular nature of electron-hole recombination.
For fluence $F$ between 10 and 100 $\mu$J/cm${}^{2}$ (for the given choice of
parameters) one indeed observes a change in the character of the
pre-bottleneck dynamics, from being governed by the electron-hole-pair
generation to being governed by the bi-molecular recombination of EHPs. 

In superconductors such a change in character of the dynamics has never been
observed at high excitation densities. Most likely this can be attributed to
the fact that the gap gets fully suppressed before such a regime can be
reached. Our data on YbB$_{12}$, however, do display such a transition,
underscoring the fact that the hybridization gap is extremely robust against
electronic excitation.

\end{document}